\documentclass[traditabstract]{aa} % for the abstract without structuration
\usepackage{graphicx}
\usepackage{txfonts}
\usepackage{supertabular}
\usepackage{textcomp}
\usepackage{booktabs}
\usepackage{lscape}
\usepackage{natbib}
\usepackage{xspace}
\usepackage{tabularx}
\usepackage{pst-all}

\graphicspath{{./}}
\usepackage{txfonts}

\def\C120{3C\,120}
\def\V16{BMT}
\def\Vy6{RoBoTT}
\def\BII{BEST\,II}

\begin{document}

\title{The stability of the optical flux variation gradient for 3C\,120}
\titlerunning{The stability of the optical flux variation gradient for 3C\,120}

\author{Michael Ramolla
  \inst{1}
  \and
  Francisco Pozo  Nu\~nez
  \inst{1}
  \and
  Christian Westhues
  \inst{1}
  \and
  Martin Haas
  \inst{1}
  \and
  Rolf Chini
  \inst{1,2}
}
\institute{Astronomisches Institut, Ruhr--Universit\"at Bochum,
  Universit\"atsstra{\ss}e 150
  , 44801 Bochum, Germany\\
  \email{ramolla@astro.rub.de}
  \and
  Instituto de Astronom\'{i}a,
  Universidad Cat\'{o}lica del Norte, Avenida Angamos 0610,
  Casilla 1280 Antofagasta, Chile
}

\date{Received --; accepted --}

\abstract{ New $B$- and $V$-band monitoring in 2014 -- 2015 reveals
  that the Seyfert 1 Galaxy, \C120, has brightened by a magnitude of
  $1.4$, compared to our campaign that took place in 2009 -- 2010.  This
  allowed us to check for the debated luminosity and time-dependent
  color variations claimed for SDSS quasars. For our \C120\ data, we
  find that the $B/V$ flux ratio of the variable component in the bright
  epoch is indistinguishable from the faint one.  We do not find any
  color variability on different timescales ranging from about $1$ to
  $1800$\,days. We suggest that the luminosity and time-dependent color
  variability is an artifact caused by analyzing the data in magnitudes
  instead of fluxes. The flux variation gradients of both epochs yield
  consistent estimates of the host galaxy contribution to our 7$\farcs$5
  aperture.  These results confirm that the optical flux variation
  gradient method works well for Seyfert galaxies. }

\maketitle

\section{Introduction}  
\label{sec:introduction}  The UV to optical color ratio of the total
  AGN, measured in magnitude units, becomes bluer as the AGN brightens
(e.g. \citealt{2011A&A...525A..37M}). Some studies attribute this change
in color to the spectral hardening of the variable component
(\citealt{1999MNRAS.306..637G,2005ApJ...633..638W,1990ApJ...354..446W,2000ApJ...540..652W}).
However, there is ample evidence to suggest that the color of the
variable component stays constant and that the corresponding flux
variation gradient is offset from the origin of the flux--flux diagram
(\citealt{1981AcA....31..293C,1992MNRAS.257..659W,1997MNRAS.292..273W,1996A&A...312...55P,2010ApJ...711..461S}).
In this case, the `bluer when brighter' total observed fluxes, which
 are measured in a finite aperture, are explained by the
superimposition of a constant red host galaxy (including
non-varying emission lines) and a varying blue AGN.  Even if the
measured host contribution is small, compared to the required offset of
the flux variation gradient (FVG), there is no significant curvature
seen in the flux--flux diagrams (\citealt{2011ApJ...731...50S}).

The constancy of the optical colors of the variable component is
physically plausible if the variable emission has a hot thermal origin,
as expected for an accretion disk (AD). In this case, fluxes in the optical range lie
on the Rayleigh-Jeans tail, and they scale almost linearly with
temperature. 

Recently, \cite{2014ApJ...792...54S} report time-dependent $g$/$r$
color variability of the SDSS Stripe 82 quasar sample. A blue color at
variations inside $<$30~days (with smaller amplitudes) gradually changes
to redder colors on larger timescales $>$1000~days (and larger
amplitudes).  Furthermore, \cite{2014ApJ...792...54S} claim that the FVG
method lacks rigor and is, therefore, not valid.  Consequently, one also
expects to find similar behavior for Seyfert galaxies, the less luminous
siblings of quasars.

Using data from 2009 -- 2010, \cite{2012A&A...545A..84P} performed
$B$-,$V$-band monitoring of the Seyfert~1 Galaxy 3C\,120\  including
dense daily observations over a period of five months.  Here we report
new $B$-,$V$-band results from a six-month monitoring project in 2014 and
2015.  The total brightening (by about 1.4 mag) and the dense
time-sampling of observations allow us to check whether or not
brightness and time-dependent color variations are present.

\section{Observations and data reduction}       
\label{sec:Observations}
The new photometric data at Johnson $B$ and $V$ was observed between 27
August 2014 and 3 March 2015 at the Universit\"atssternwarte Bochum,
near Cerro Armazones.  We combined the \Vy6\ telescope
data (\citealt{2012A&A...545A..84P}) with new data from the \BII\
(\citealt{2009A&A...506..569K}).

\begin{figure}
  \centering
  \includegraphics[width=.9\columnwidth]{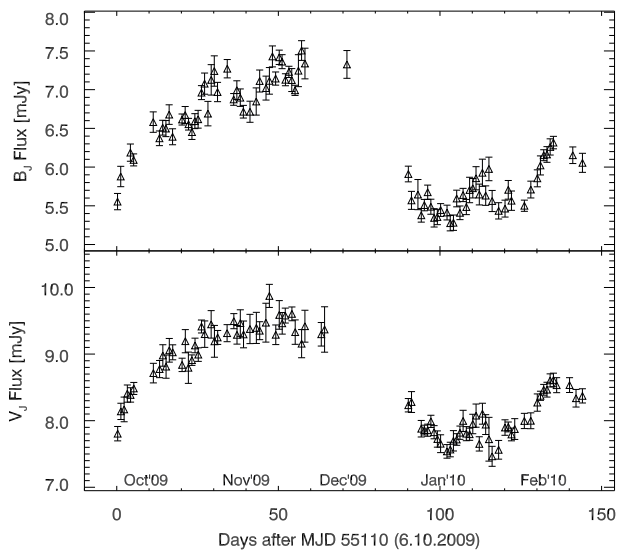}
  \caption{Lightcurves in the 2009 -- 2010
    epoch obtained with the \Vy6\ telescope.
    All fluxes are corrected for
    galactic foreground extinction. }
  \label{fig:lc_2010}
\end{figure}

\begin{figure}
  \centering
  \includegraphics[width=.9\columnwidth]{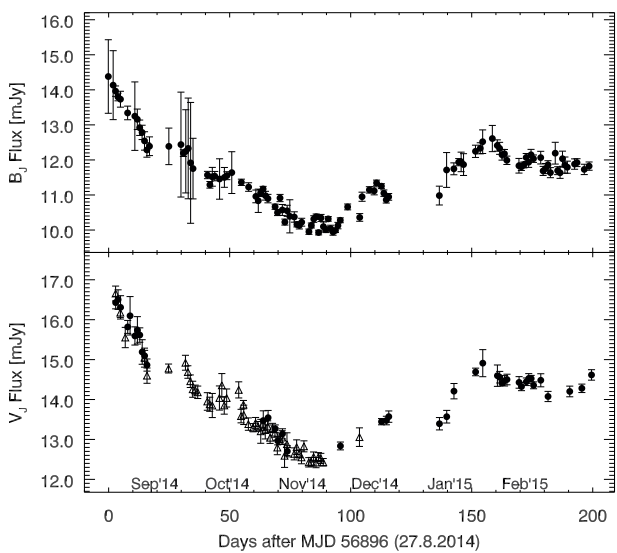}
  \caption{Combined lightcurves in the 2014 -- 2015
    epoch, obtained with two different telescopes.
    Filled circles correspond to \BII, while open triangles
    represent the \Vy6\ observations. All fluxes are corrected for
    galactic foreground extinction. }
  \label{fig:lc_2014}
\end{figure}

All data has been corrected for the latest revision of galactic
foreground extinction\footnote{$A_{\lambda}^{B} = 1.079$,
  $A_{\lambda}^{V} = 0.816$} by \cite{2011ApJ...737..103S}
and the corresponding lightcurves are displayed in Figs.\ref{fig:lc_2010} and~\ref{fig:lc_2014}. 

\section{Results}
\label{sec:results}
\subsection{Flux-flux diagrams} 
\label{ssec:host_lum}
\begin{figure}
  \centering
  \includegraphics[width=.95\columnwidth,clip=true]{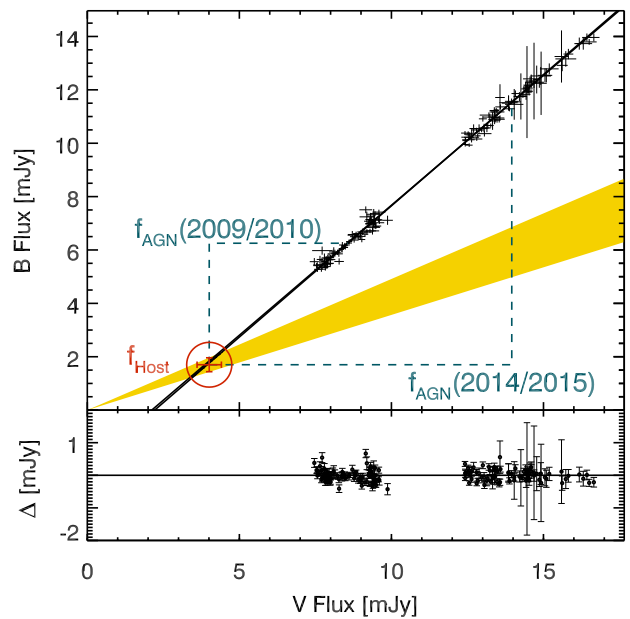}
  \caption{ $B$ versus $V$ flux variations of \C120\ in the $7\farcs5$
    aperture. Each measurement is drawn as a thin cross in which the line
    length corresponds to the photometric uncertainties in the
    respective filters.
    The yellow filled area is the assumed host color
    $\Gamma_{BV, \rm Host}$, drawn to cover $B=(1.70\pm0.27)$\,mJy and
    $V=(4.01\pm0.62)$\,mJy host fluxes, as determined by
    \cite{2010ApJ...711..461S} in an $8\farcs3$ aperture.  Our host
    flux estimate is the datapoint  in the red circle.  All data from 2009
    until 2015 was used to determine the AGN slope.
    The black continuous line covers the upper and lower standard
    deviations in the AGN slope, given by the OLS bisector fit.
}
  \label{fig:fvg_bv_all}
\end{figure}
Total $B$- versus $V$-band fluxes for the two epochs are shown in
Fig.~\ref{fig:fvg_bv_all}, with one that had low luminosity in
2009--2010 and one that had high luminosity in 2014--2015. Between the
two epochs, the AGN luminosity increased by a magnitude of about 1.4 in
both filters.  The slope of AGN variability is well matched by a linear
relation with $\Gamma_{BV} = 0.979\pm0.005$.

We consider the host colors of \C120\ by \cite{2010ApJ...711..461S} in a
$8\farcs3$ aperture, corrected for galactic foreground extinction
(\citealt{2011ApJ...737..103S}).  Assuming that these will be similar
when looked at in our $7\farcs5$ aperture, this allows us to compute our
own host fluxes by measuring the center of gravity of the area that is
encased by the cone of AGN slope $\Gamma_{BV}$-fitting uncertainties and
Sakata's host color $\Gamma_{BV, \rm Host}$ (inside the red circle of
Fig.~\ref{fig:fvg_bv_all}). The results for the $B$- and $V$-band host
fluxes (Table \ref{tab:fluxes}) agree well with the values of Sakata. We
also fit the slopes of the individual epochs of 2009-2010 and 2014-2014
separately. For both epochs the slope is slightly flatter than the
combined one, but they agree within the errors.

The offset of the combined slope ($2009 - 2015$) to the individual epoch
slopes could imply that the short-term variability is redder than the
long-term one.  A more likely physical explanation, however, is based on
the systematically different instrumental point spread functions (PSF).
The PSF of the \Vy6\ is slightly larger than that of the \BII.  The host
also has a different spatial flux distribution at $B$ and $V$.  This
leads to a larger host galaxy contribution in the $2009 - 2010$ light
curves, and the host galaxy contribution is larger in $V$ than in $B$.
In the net effect, compared to the $2014 - 2015$ data, the $2009 - 2010$
data appear slightly shifted in the FVG diagram toward the right,
resulting in a marginally steeper overall $\Gamma_{BV}$.  We believe
that these negligible effects do not alter the basic results and
conclusions on the stability of the FVG method because for all three
data sets ($2009 - 2010$, $2014 - 2015$, and $2009 - 2015$) the host
galaxy contribution is in excellent agreement within the errors.  In $B$-band we measure a minimum $f_{B, \rm AGN} = 3.59$\,mJy and a maximum of
$f_{B, \rm AGN} = 12.69$\,mJy.  Correspondingly, the increase in $V$
ranges from a minimum of $f_{V, \rm AGN} = 3.57$\,mJy to a maximum of
$f_{V, \rm AGN} = 12.76$\,mJy.

\begin{table}
\begin{center}
 \hfill{}
 \caption{Results of the FVG analysis, separated into epochs and filter
   sets $B$ and $V$. All fluxes in mJy are corrected for galactic
   foreground extinction as determined by
   \cite{2011ApJ...737..103S}. The host fluxes may also contain
   contributions from non-variable, narrow and broad emission lines.  }
\label{tab:fluxes}
\begin{tabular}{@{}lccc}
\hline
\hline
Epoch, Filter& $f_{\rm Host}$ & $f_{\rm AGN}$ & $\Gamma_{\rm AGN}$  \\
\cmidrule{2-3}
 & \multicolumn{2}{c}{(mJy)}&  \\
\hline
2009\,--\,2015, $B$ &  $1.69 \pm 0.28$ & --                 &  $0.979 \pm 0.005$  \\
2009\,--\,2015, $V$ &  $3.89 \pm 0.29$ & --                 & ``    \\
2009\,--\,2010, $B$ &  $1.59 \pm 0.26$ & $4.66 \pm 0.72$    &  $0.963 \pm 0.036$ \\
2009\,--\,2010, $V$ & $3.72 \pm 0.32$  & $4.84 \pm 0.76$    & ``    \\
2014\,--\,2015, $B$ &  $1.51 \pm 0.26$ & $10.03 \pm 1.08$   & $0.959 \pm 0.015 $ \\
2014\,--\,2015, $V$ & $3.48 \pm 0.31$  & $10.47 \pm 1.14$   & ``    \\
\end{tabular}
\end{center}
\end{table}

\subsection{Search for timescale-dependent AGN colors} 
\label{ssec:color_var}
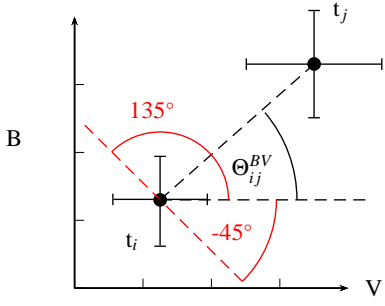
\begin{figure}
  \centering
{
\psscalebox{.90}{
\begin{pspicture}(0,-2.8660364)(5.8670835,2.8660364)
\rput(1.0,-2.001047){\psaxes[linecolor=black, linewidth=0.02, tickstyle=full, tickstyle=top, axesstyle=axes, labels=none, ticks=all, dx=1.0cm, dy=1.0cm]{->}(0,0)(0,0)(4,4)}
\rput[bl](0.0,0.0988407){B}
\rput[r](5.5,-2.){V}
\psdots[linecolor=black,  dotsize=0.2](2.25,-0.69988406)
\psdots[linecolor=black, dotsize=0.2](4.5,1.300116)
\psline[linecolor=black, linewidth=0.02, linestyle=dashed, dash=0.17638889cm 0.10583334cm](2.25,-0.69988406)(5.25,-0.69988406)
\psline[linecolor=black, linewidth=0.02, linestyle=dashed](2.25,-0.69988406)(4.5,1.300116)(4.5,1.300116)
\psarc[linecolor=black, linewidth=0.02, dimen=outer](2.25,-0.69988406){2.0}{0.0}{40.0}
\psline[linewidth=0.02,linecolor=red, linestyle=dashed](1.15,0.4)(3.45,-1.90)
\psarc[linecolor=red, linewidth=0.02, dimen=outer](2.25,-0.69988406){1.7}{-45.0}{0.0}
\psarc[linecolor=red, linewidth=0.02, dimen=outer](2.25,-0.69988406){1.}{.0}{135.0}
\rput[bl](3.,-1.30){\color{red}-45$^{\circ}$}
\rput[bl](1.8,0.50){\color{red}135$^{\circ}$}
\rput[bl](3.3,-0.5){$\Theta_{ij}^{BV} $}
\rput[bl](1.75,-1.45){t$_i$}
\rput[r](5.0,2.0501158){t$_j$}
\psline[linecolor=black, linewidth=0.02, tbarsize=0.07cm 5.0]{|-|}(1.55,-0.692961)(2.95,-0.692961)
\psline[linecolor=black, linewidth=0.02, tbarsize=0.07cm 5.0]{|-|}(2.25,-1.392961)(2.25,-0.0492961)
\psline[linecolor=black, linewidth=0.02, tbarsize=0.07cm 5.0]{|-|}(3.5,1.30)(5.5,1.3)
\psline[linecolor=black, linewidth=0.02, tbarsize=0.07cm 5.0]{|-|}(4.5,.5)(4.5,2.1)

\end{pspicture}
}
}%psscalebox
\caption{Scheme of the color slope $\Theta_{ij}^{BV}$ in the $B$- to
  $V$-band flux plane between two measurements $t_i$ and $t_j$ with
  photometric uncertainties. If $\Theta$ falls below $-45^{\circ}$ or
  above $135^{\circ}$ we add or subtract $180^{\circ}$
  respectively. Restricting ourselves to the half circle, marked by the
  red arcs, ensures proper averaging of $\Theta_{ij}^{BV}$ for low
  values of $\tau$, where photometric uncertainties dominate the slopes. A
  restriction to the full circle, or $-90^{\circ}$ to $90^{\circ}$, would
  bias the slope towards $0^{\circ}$ if the photometric errors are large
  compared to the true change of flux. }
  \label{fig:angles}
\end{figure}

\begin{figure}
  \centering
  \includegraphics[width=.9\columnwidth,clip=true]{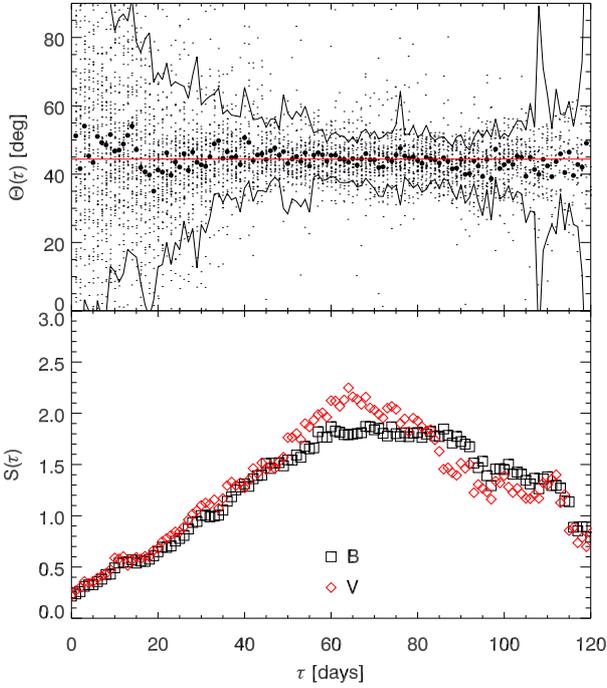}
  \caption{Top: Small dots represent $\Theta_{ij}^{BV} $ between all
    possible permutations of measurements $i,j$ with $\tau = |t_j -
    t_i|$. Larger dots mark the averages inside a bin of one day
    with their uncertainty plotted as a continuous black line. The red
    straight line at $44.4^{\circ}$ corresponds to the linear fit of all
    data $\Gamma_{BV} = 0.979$.  Bottom: The structure function $S$ of
    the variability; black boxes for $B$-band and red diamonds for $V$-band. }
  \label{fig:short_angles}
\end{figure}

\begin{figure}
  \centering
  \includegraphics[width=.9\columnwidth,clip=true]{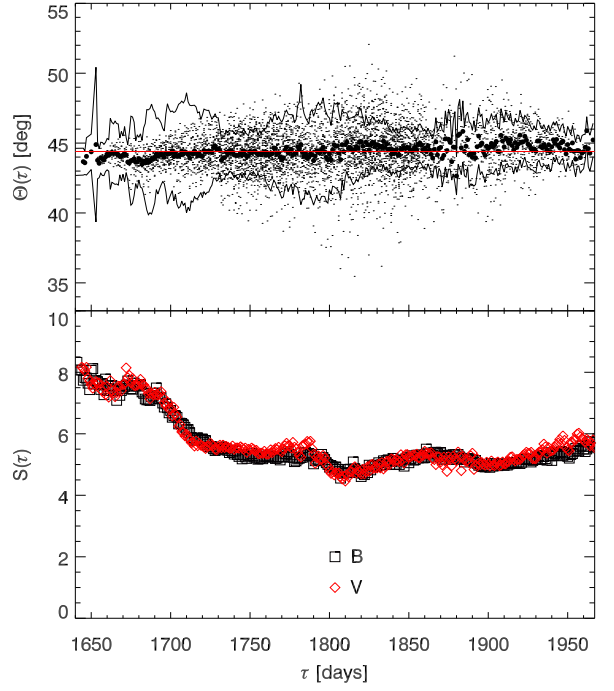}
  \caption{Same as Fig. \ref{fig:short_angles}, here for the high
    values of $\tau$ between the epochs of 2009 -- 2010 and 2014 -- 2015. }
  \label{fig:long_angles}
\end{figure}

The aim of this section is to investigate potential, timescale-dependent
variability that has recently been observed by
\cite{2014ApJ...792...54S} in a sample of Stripe 82 SDSS quasars. The
ensemble color variability of their quasars was not constant for all
observed redshifts, showing rather blue slopes on short timescales of $<
30$\,days that turn to redder slopes on longer timescales.

Here we improve their approach in some aspects. Instead of using
magnitudes, we make use of fluxes $f_{\nu}$ directly. As pointed out by
\cite{2014ApJ...783...46K} and references therein, fitting a straight
line in magnitude--magnitude space (e.g. \citealt{2014ApJ...792...54S})
relies heavily on the contamination of the baseline flux by host
galaxies and emission lines.

The color $\Theta_{ij}^{BV} $ of an arbitrary flux variation in $B$- and
$V$-band in the time interval $\tau = | t_j - t_i |$ is computed using
Equation \ref{eq:theta}. The geometry is explained in
Fig.~\ref{fig:angles}.

\begin{equation}
  \label{eq:theta}
  \Theta_{ij}^{BV} (\tau) = arctan \left( \frac{f_B(t_j) - f_B(t_i)}{f_V(t_j) - f_V(t_i)}\right) % + 45^{\circ}
.\end{equation}

Like \cite{2014ApJ...792...54S}, we restrict ourselves to the range of
slopes $-45^{\circ}<\Theta_{ij}^{BV}<135^{\circ}$. The slopes on the
other half circle represent flux decreases and are rotated by $180^{\circ}$ to their
  equivalent slopes for a flux increase. The
offset of $45^{\circ}$ is chosen to equally weigh cases of both bluer
(i) when brighter ($45^{\circ}<\Theta_{ij}^{BV}<135^{\circ} $), and (ii)
redder, when brighter ($-45^{\circ}<\Theta_{ij}^{BV}<45^{\circ} $),
slopes in the averaging process. The angle of $45^{\circ}$ represents no
color change.

Our photometric measurements in all observed epochs offer
$\Theta_{ij}^{BV}$ for many combinations of $t_i$ and $t_j$, with the
lowest sampling interval of $\tau = 1$\,day. It is useful to average all
$\Theta_{ij}^{BV}$ inside bins with this size. The average of these bins
is described by

\begin{equation}
  \label{eq:theta_avg}
  \overline{\Theta}^{BV} (\tau) = \frac{1}{N} \sum_{ij}^{N}{\Theta_{ij}^{BV} (\tau)}
.\end{equation}

The results of this approach for the $B$- and $V$-band fluxes are shown
in Fig.~\ref{fig:short_angles}, together with the structure function
$S(\tau),$ as defined by Equation \ref{eq:sf}. The latter is an
indicator of the strength of the variability on a specific time scale
$\tau$. The black solid lines enclose the average photometric
uncertainty of each $\tau$ bin.

\begin{equation}
  \label{eq:sf}
  S(\tau) = \sqrt{ \frac{1}{N}  \sum_{ij}^{N}{(f_{\nu}(t_j) - f_{\nu}(t_i))^2} }
.\end{equation}

Analyzing photometric fluxes with large uncertainties
$\sigma_i,\sigma_j$, compared to the difference in fluxes with
$|f_{\nu}(t_j)~-~f_{\nu}(t_i)|~<~\sqrt{\sigma_i^2+\sigma_j^2}$ , can
cause a bias on $\Theta_{ij}$. We assume a case of flux that is constant
except for noise that is described by different photometric
uncertainties for $B$- and $V$-band with ${\sigma_{fB}} >>
{\sigma_{fV}}$. Then, referring to the sketch in Fig.~\ref{fig:angles},
we have $\Delta f_B >> \Delta f_v$ for samples of measurements drawn
from these distributions. As a result, we obtain
$\overline{\Theta}(\tau) \approx 90^{\circ}$.

The structure function $S(\tau)$ for the short timescales in
Fig.~\ref{fig:short_angles} shows that for $\tau < 20$\,days the
variations are still on the order of typical photometric
errors. As a consequence, the distribution of $\Theta_{ij}^{BV}$ is very
broad in this region. The shape of the structure function is very
similar for both bands with a slightly larger variation in amplitudes
around $60$ days for the $V$-band. For comparison purposes, we plot the
previous linear fit result of $\Gamma_{BV} = 0.979$ as solid red
line. In this early epoch, the majority of slopes
$\overline{\Theta}^{BV} (\tau)$ is about $3^{\circ}$ higher than the
linear approximation but with only low significance and large
scatter. From $20$ to $90$\,days, the scatter decreases and the average
cannot be distinguished from the linear fit. A potential contribution of
$\sim 5\%$ H$\beta$ at $25$\,days (\citealt{2012A&A...545A..84P}) may
introduce additional flux to the $V$-band and shifts the slope downward
in this region. However, there is no significant trace of this
effect. After $90$\, days, the number of data points in the $\tau$ bins
decreases naturally, causing larger scatter and a minimally lower
average slope of $42^{\circ}$. In summary, the plot reveals no clear
difference in $\overline{\Theta}^{BV}$ on timescales $\tau$ within a
single observation epoch of \C120.

In Figure \ref{fig:long_angles}, we restrict the plot to $1640 < \tau <
1960$, so that all $\Theta_{ij}^{BV}$ have one data point in our early
epoch of 2009 -- 2010 and one in the late epoch of 2014 -- 2015. In
total, all angles $\overline{\Theta}^{BV}$ are consistent with our
linear fit result, taking the photometric uncertainties into
consideration.  Within the errors, there appears a positive slope from
$~44^{\circ}$ to $~45^{\circ}$ for the considered range of $\tau$. This
is, however, consistent with a potential host flux offset between the
two epochs that originates from different PSFs as explained in
Section~\ref{ssec:host_lum}.

\section{Discussion and conclusions}
\label{ssec:discussion}

Our results show a highly linear evolution of $B$ versus $V$ fluxes for
\C120\ on all timescales -- from days to several years. Additionally,
the color is highly constant, despite the AGN undergoing a strong
increase of $1.4$ mag in flux over a duration of five years.

Calculating their time dependent colors $\theta(\tau)$ in
magnitude--magnitude space, \cite{2014ApJ...792...54S} report a
significant change of color for $g$- and $r$-band data of a large SDSS
quasar sample.  The colors change from short timescale (small amplitude)
blue colors to long timescale (large amplitude) red colors. Considering
a total observed flux $f(t) = f_{\rm Host} + f_{\rm AGN}(t)$, composed
of constant host galaxy component and variable AGN, the difference in
magnitude space is

\begin{equation}
  \label{eq:sun}
  \Delta m_{ij}(\tau) =2.5\, log\left( \frac{f_{\rm AGN}(t_j) +  f_{\rm Host}}{f_{\rm AGN}(t_i) +  f_{\rm Host}} \right)
.\end{equation}

This means that a constant host causes a reduction in the observed
magnitude difference. Because the host galaxy SED is red, compared to
that of the AGN, the magnitude difference will be reduced {more} for the
red $\Delta m_r(\tau)$ compared to the blue $\Delta m_g(\tau)$ and
therefore $\theta(\tau) = arctan (\Delta m_r(\tau)/\Delta m_g(\tau) )$
turns to artificially blue colors. Since the structure function of the
SDSS quasars rises towards long timescales
(\citealt{2014ApJ...792...54S}), we simultaneously observe large
amplitudes of flux variations. Because the host galaxy contribution is
reduced in such cases, $\theta(\tau)$ will be of a redder color.  In our
flux--flux approach, however, this issue is completely avoided.
Therefore, we suggest that the luminosity and time-dependent color
variability observed by Sun et al. is an artifact of data analysis in
magnitudes instead of fluxes.

The intrinsic brightness change of any astronomical object may be caused
by a change in the area and/or the temperature $T$ of the emitting
surface. While an increase in area would simply cause a constant color,
the temperature change always has a degree of curvature. Only in the
Rayleigh-Jeans approximation is the flux again proportional to the
temperature $T$. Using the blue side $\lambda$ of the Johnson $B$-band
in $ h\, c < \lambda\, k_B\, T$, the temperature $T$ must be higher than
36,000\,K, because it is at the lower end of typical effective AD
temperatures. The existence of a linear relationship between two
(optical) fluxes shows that the variable component is hot enough to be
approximated by Rayleigh Jeans.

Another mechanism for a brightness change is variable extinction. In
this case, the extinction law influences the color of the varying
component, with the bluer band usually more affected than the red band.
As a result, extinction can provide a different color slope than a
temperature fluctuation in the AD.  However, extinction variations
resulting from dust clouds that move into the line of sight are random
events and should occur with diverse amplitude and timescales. Because
they are independent of the variations in the AD with different its
color, non-linear behavior of the total fluxes should then be observed
in individual objects. Averaging a large sample of AGN flux variations,
such random disturbances of the slope in the flux--flux diagrams will be
washed out.

Our results are consistent with variations that stem from temperature
changes in the AD, where the $B$- and $V$-bands are placed in the
Rayleigh-Jeans tail of the thermal emission. There is no evidence for a
short timescale, bluer color variation that may be caused by variable
extinction in the line of sight.

In summary, our finding makes a good case for the use of the rest-frame
optical FVG method on different timescales of low luminosity AGN. To
further confirm this result, more longterm observations of varying
Seyferts (and also quasars) with high photometric precision and dense
(daily) temporal sampling will be required.

\begin{acknowledgements}
  This research made use of the NASA/IPAC Extragalactic Database (NED)
  which is operated by the Jet Propulsion Laboratory, California
  Institute of Technology, under contract with the National Aeronautics
  and Space Administration (NASA).  This publication is supported as a
  project of the Nordrhein-Westf\"alische Akademie der Wissenschaften
  und der K\"unste, in the framework of the academy program of the
  Federal Republic of Germany and the state of Nordrhein-Westfalen. This
  work is supported by the DFG Program (HA 3555/12-1).  The observations
  on Cerro Armazones benefited from the support of guardians, Hector
  Labra, Gerardo Pino, Roberto Munoz, and Francisco Arraya. We thank
    the referee, Ian Glass, for his helpful comments and careful review
    of the manuscript.
\end{acknowledgements}

%\bibstyle{aa} % style aa.bst

\bibliographystyle{aa} % style aa.bst
\bibliography{rm_aanda_compatible}%\usepackage{subfig}

\begin{appendix}
\section{Measured fluxes}
\label{sec:measured_fluxes}
\vspace{1cm}
\tablefirsthead{%
\hline
MJD & Flux  & Error \\ 
\cmidrule(r){2-3}
days&\multicolumn{2}{c}{mJy} \\
\hline
}
\tablehead{%
\multicolumn{2}{l}{Table \ref{table1} continued.}  & \\
\hline
MJD & Flux  & Error  \\
\hline
}
\tablelasttail{%
\hline
}
\tablecaption[]{Galactic foreground extinction corrected $B$-band fluxes from 2009 until 2015. 
 }
  \label{table1}
\begin{supertabular}{@{\hspace{12mm}}l@{\hspace{12mm}}c@{\hspace{12mm}}c@{\hspace{12mm}}}
55110.237&5.55421&0.105191     \\
55111.234&5.8784&0.131488      \\
55114.225&6.18674&0.111323     \\
55115.223&6.09283&0.077608     \\
55121.283&6.58088&0.132027     \\
55123.203&6.37179&0.0947509    \\
55124.209&6.5088&0.0957203     \\
55125.209&6.49741&0.103942     \\
55126.209&6.67802&0.126342     \\
55127.264&6.39158&0.103038     \\
55130.204&6.61423&0.0713771    \\
55131.166&6.67221&0.110278     \\
55132.208&6.55387&0.075334     \\
55133.260&6.45186&0.0950582    \\
55134.166&6.59332&0.0966424    \\
55135.172&6.61771&0.114972     \\
55136.209&6.96171&0.0903407    \\
55137.175&7.07786&0.136614     \\
55138.209&6.69046&0.158005     \\
55139.218&7.12651&0.196419     \\
55140.175&7.24073&0.196107     \\
55141.175&6.96943&0.124644     \\
55144.185&7.27056&0.118578     \\
55146.175&6.874&0.0758508      \\
55147.153&6.99352&0.121185     \\
55148.216&6.89657&0.112869     \\
55149.175&6.71507&0.0817101    \\
55151.222&6.71472&0.136257     \\
55153.145&6.84699&0.175694     \\
55154.217&7.11164&0.142083     \\
55156.181&7.02015&0.15446      \\
55157.226&7.11348&0.171963     \\
55158.164&7.42939&0.135409     \\
55159.148&7.14331&0.085071     \\
55160.268&7.41794&0.0926098    \\
55161.144&7.36146&0.0904423    \\
55162.188&7.12791&0.0774404    \\
55163.262&7.22946&0.0724561    \\
55164.148&7.12385&0.148103     \\
55165.159&6.99121&0.065618     \\
55166.200&7.24529&0.206256     \\
55167.177&7.50535&0.127324     \\
55168.226&7.3375&0.197946      \\
55181.213&7.32589&0.179363     \\
55200.209&5.91065&0.101208     \\
55201.179&5.57005&0.116409     \\
55203.179&5.64471&0.194746     \\
55204.216&5.37734&0.088279     \\
55205.160&5.50883&0.0735336    \\
55206.194&5.67506&0.0920804    \\
55207.180&5.48779&0.0972899    \\
55208.188&5.34417&0.118273     \\
55209.185&5.35524&0.0988943    \\
55210.200&5.44533&0.0826862    \\
55212.181&5.41264&0.095357     \\
55213.176&5.27572&0.0994904    \\
55214.169&5.28162&0.0990277    \\
55215.171&5.59415&0.107598     \\
55216.185&5.40936&0.0881853    \\
55217.145&5.6395&0.0854541     \\
55218.158&5.48408&0.0979306    \\
55219.161&5.70425&0.164234     \\
55220.156&5.72685&0.125466     \\
55221.153&5.8559&0.149499      \\
55222.168&5.64563&0.134439     \\
55223.128&5.92687&0.175566     \\
55224.146&5.63153&0.125795     \\
55225.142&5.97617&0.152117     \\
55226.138&5.56186&0.13634      \\
55228.152&5.43482&0.105249     \\
55230.126&5.4637&0.10911       \\
55231.127&5.70627&0.120521     \\
55232.122&5.56605&0.124716     \\
55236.111&5.50081&0.0723353    \\
55238.111&5.70972&0.108597     \\
55240.117&5.85487&0.10974      \\
55241.081&6.02139&0.122924     \\
55242.096&6.15323&0.0706417    \\
55243.097&6.16286&0.101422     \\
55244.087&6.26022&0.104736     \\
55245.096&6.31632&0.0806359    \\
55251.089&6.15306&0.107161     \\
55254.081&6.05505&0.123017     \\
56895.368&14.3804&1.04936      \\
56897.366&14.1336&0.983937     \\
56898.358&13.9633&0.147968     \\
56899.358&13.7959&0.109101     \\
56900.385&13.7324&0.226593     \\
56903.344&13.3408&0.194499     \\
56906.327&13.2489&0.976908     \\
56907.343&13.1597&0.291964     \\
56908.330&12.9185&0.220106     \\
56909.325&12.7856&0.204544     \\
56910.333&12.5424&0.262686     \\
56911.312&12.2846&0.206587     \\
56912.377&12.3907&0.266982     \\
56920.377&12.3876&0.521437     \\
56925.375&12.4381&1.49668      \\
56926.373&12.1945&0.0961339    \\
56927.369&12.2455&1.18698      \\
56928.367&12.3288&1.43678      \\
56929.364&11.9162&1.7241       \\
56930.361&11.7529&0.864746     \\
56936.336&11.5692&0.10415      \\
56937.335&11.29&0.0942527      \\
56938.334&11.5188&0.268483     \\
56939.334&11.5381&0.133113     \\
56941.330&11.4562&0.576352     \\
56943.325&11.5018&0.283021     \\
56944.352&11.5507&0.11299      \\
56946.374&11.6383&0.596045     \\
56950.347&11.3615&0.0897234    \\
56953.332&11.2265&0.121999     \\
56956.323&10.9593&0.181796     \\
56957.321&10.8346&0.332822     \\
56958.321&11.0549&0.0997387    \\
56959.313&11.1725&0.0734273    \\
56960.311&10.9795&0.125564     \\
56961.314&10.8968&0.120839     \\
56964.273&10.665&0.0736191     \\
56965.272&10.496&0.084032      \\
56966.310&10.913&0.103403      \\
56967.265&10.5722&0.156577     \\
56968.306&10.2267&0.0856314    \\
56969.261&10.5487&0.157831     \\
56970.352&10.391&0.472495      \\
56972.202&10.3721&0.148061     \\
56973.287&10.1588&0.0703249    \\
56974.284&10.1296&0.0993075    \\
56975.258&10.2233&0.112217     \\
56978.251&9.95252&0.0759752    \\
56979.248&10.129&0.0708522     \\
56980.246&10.3133&0.0943685    \\
56981.243&10.3805&0.0763472    \\
56982.240&9.92794&0.066818     \\
56983.237&10.3451&0.0996063    \\
56984.235&10.098&0.130562      \\
56985.231&10.0103&0.0832779    \\
56986.222&10.3187&0.0663072    \\
56987.215&10.0418&0.095233     \\
56988.211&9.94455&0.0923461    \\
56989.211&9.98914&0.0576447    \\
56990.230&10.1267&0.0918454    \\
56991.199&10.2779&0.0700807    \\
56994.243&10.6593&0.083241     \\
56999.226&10.3593&0.109408     \\
57000.196&10.9499&0.13015      \\
57003.220&11.1454&0.090543     \\
57005.216&11.1176&0.114169     \\
57006.198&11.3441&0.0799838    \\
57008.191&11.2517&0.069811     \\
57009.205&11.0519&0.0910484    \\
57010.192&10.8618&0.0919861    \\
57011.188&10.946&0.0958581     \\
57032.126&10.9831&0.268319     \\
57035.126&11.7138&0.496919     \\
57038.123&11.7465&0.169581     \\
57040.123&11.9359&0.102771     \\
57041.122&11.9438&0.278621     \\
57042.122&11.8676&0.310776     \\
57047.059&12.2454&0.174759     \\
57049.023&12.3322&0.15833      \\
57050.022&12.5201&0.336703     \\
57054.020&12.6104&0.374115     \\
57056.067&12.4112&0.161701     \\
57057.020&12.3306&0.171476     \\
57058.020&12.1477&0.15279      \\
57059.019&12.1763&0.122279     \\
57060.019&11.9898&0.11905      \\
57065.016&11.8339&0.208689     \\
57066.016&11.8045&0.104941     \\
57067.052&11.8566&0.0844698    \\
57068.014&12.0699&0.114009     \\
57069.014&11.9332&0.14584      \\
57070.013&12.1462&0.155033     \\
57071.013&12.0355&0.106119     \\
57074.011&12.0657&0.180716     \\
57075.010&11.6865&0.129328     \\
57076.010&11.7541&0.144578     \\
57077.009&11.861&0.11675       \\
57078.008&11.6374&0.143872     \\
57080.008&12.1926&0.310672     \\
57081.007&11.6933&0.112834     \\
57082.006&11.6342&0.160317     \\
57083.005&12.0345&0.216004     \\
57084.004&11.8517&0.231508     \\
57085.004&11.7921&0.172509     \\
57088.002&11.8821&0.148556     \\
57089.002&11.9226&0.112329     \\
57092.004&11.7295&0.147702     \\
57093.997&11.8236&0.117837     \\
\end{supertabular}             
                         
\clearpage
\tablefirsthead{%
\hline
MJD & Flux  & Error \\
\cmidrule(r){2-3}
days&\multicolumn{2}{c}{mJy} \\
\hline
}
\tablehead{%
\multicolumn{2}{l}{Table \ref{table2} continued.}  & \\
\hline
MJD & Flux  & Error  \\
\hline
}
\tablelasttail{%
\hline
}
\tablecaption[]{Galactic foreground extinction corrected $V$-band fluxes from 2009 until 2015.
 }
  \label{table2}
\begin{supertabular}{@{\hspace{12mm}}l@{\hspace{12mm}}c@{\hspace{12mm}}c@{\hspace{12mm}}}
55110.283& 7.80492& 0.107975 \\
55111.280& 8.13988& 0.122128 \\
55112.281& 8.16902& 0.185274 \\
55113.274& 8.40669& 0.129154 \\
55114.271& 8.38676& 0.107963 \\
55115.269& 8.48413& 0.0902408\\
55121.329& 8.71428& 0.14702  \\
55123.249& 8.77489& 0.128866 \\
55124.255& 8.97913& 0.161263 \\
55125.255& 8.8118 &0.174615  \\
55126.255& 9.05692& 0.177551 \\
55127.311& 9.02664& 0.112698 \\
55130.215& 8.83791& 0.100329 \\
55131.213& 9.19359& 0.173977 \\
55132.161& 8.79508& 0.234881 \\
55133.248& 8.90878& 0.0787138\\
55134.213& 9.12944& 0.109715 \\
55135.219& 8.99221& 0.101696 \\
55136.219& 9.41409& 0.0970858\\
55137.222& 9.29944& 0.19867  \\
55139.208& 9.44626& 0.206467 \\
55140.222& 9.19151& 0.238798 \\
55141.221& 9.24981& 0.104757 \\
55144.138& 9.31401& 0.1281   \\
55146.223& 9.49443& 0.113287 \\
55147.199& 9.28943& 0.139888 \\
55148.227& 9.46577& 0.199261 \\
55149.223& 9.2952 &0.201076  \\
55151.211& 9.37694& 0.219707 \\
55153.192& 9.39334& 0.233614 \\
55154.228& 9.35126& 0.134791 \\
55156.135& 9.47155& 0.291827 \\
55157.216& 9.87356& 0.173444 \\
55159.196& 9.29006& 0.147111 \\
55160.280& 9.58813& 0.214793 \\
55161.192& 9.46201& 0.161921 \\
55162.139& 9.57828& 0.112009 \\
55164.196& 9.60405& 0.103562 \\
55165.207& 9.32988& 0.183195 \\
55167.188& 9.15643& 0.211352 \\
55168.214& 9.41859& 0.239692 \\
55173.276& 9.29753& 0.176752 \\
55174.295& 9.36956& 0.341082 \\
55200.199& 8.23195& 0.100521 \\
55201.189& 8.28018& 0.156595 \\
55204.206& 7.87946& 0.124108 \\
55205.170& 7.85903& 0.0732903\\
55206.184& 7.8568 &0.0907966 \\
55207.190& 7.9896 &0.0916827 \\
55208.178& 7.82703& 0.0989102\\
55209.196& 7.72464& 0.122169 \\
55210.190& 7.65428& 0.133864 \\
55212.170& 7.53899& 0.102164 \\
55213.186& 7.56678& 0.109318 \\
55214.159& 7.698 0&.152955   \\
55215.181& 7.71733& 0.0871391\\
55216.174& 7.81947& 0.0996844\\
55217.155& 7.99646& 0.159476 \\
55218.148& 7.80512& 0.105828 \\
55219.171& 7.79658& 0.0913465\\
55220.146& 7.95024& 0.146176 \\
55221.164& 8.0759 &0.174263  \\
55222.158& 7.65141& 0.107348 \\
55223.138& 8.10472& 0.157151 \\
55224.136& 7.94129& 0.148864 \\
55225.153& 7.72148& 0.328673 \\
55226.128& 7.46477& 0.150035 \\
55228.142& 7.56282& 0.140573 \\
55230.116& 7.89702& 0.116355 \\
55231.137& 7.90614& 0.0739646\\
55232.112& 7.79313& 0.0957212\\
55233.130& 7.87441& 0.156332 \\
55236.101& 7.99585& 0.114276 \\
55238.101& 7.99872& 0.119167 \\
55240.107& 8.27189& 0.130001 \\
55241.091& 8.37387& 0.0684091\\
55242.085& 8.46378& 0.0811344\\
55243.107& 8.46584& 0.1109   \\
55244.077& 8.60484& 0.0984047\\
55245.107& 8.61162& 0.10168  \\
55246.096& 8.53311& 0.112621 \\
55250.085& 8.53457& 0.111314 \\
55252.090& 8.33699& 0.131954 \\
55254.071& 8.37249& 0.106794 \\
56898.378& 16.6557& 0.186138 \\
56900.322& 16.1771& 0.120272 \\
56902.367& 15.5529& 0.254201 \\
56910.310& 15.0494& 0.238098 \\
56911.354& 14.5983& 0.191545 \\
56920.261& 14.7768& 0.112215 \\
56927.240& 14.9211& 0.186894 \\
56928.240& 14.6891& 0.156857 \\
56929.240& 14.4583& 0.157344 \\
56930.230& 14.2573& 0.200672 \\
56931.262& 14.2388& 0.121795 \\
56932.304& 14.1808& 0.155984 \\
56936.268& 13.9515& 0.220356 \\
56937.268& 13.8814& 0.177064 \\
56938.323& 13.8499& 0.301957 \\
56941.267& 14.0377& 0.302282 \\
56942.268& 14.3559& 0.294361 \\
56943.261& 13.8518& 0.218258 \\
56944.261& 14.0379& 0.226156 \\
56949.247& 14.2412& 0.20211  \\
56950.210& 13.5852& 0.169376 \\
56951.199& 13.8724& 0.104292 \\
56951.308& 13.6163& 0.242234 \\
56953.227& 13.3842& 0.155249 \\
56955.206& 13.2942& 0.103009 \\
56956.206& 13.4105& 0.140081 \\
56957.206& 13.3262& 0.122764 \\
56958.342& 13.2041& 0.300121 \\
56959.205& 13.3902& 0.108892 \\
56960.206& 13.2339& 0.221935 \\
56961.206& 13.323 &0.142973  \\
56962.206& 13.0426& 0.14191  \\
56963.220& 13.288 &0.11486   \\
56964.205& 13.1691& 0.11433  \\
56965.192& 12.7974& 0.180278 \\
56966.192& 13.0069& 0.130979 \\
56967.175& 13.0466& 0.172496 \\
56968.175& 12.5915& 0.292129 \\
56969.175& 12.8817& 0.28501  \\
56972.175& 12.6383& 0.150737 \\
56973.175& 12.8066& 0.183076 \\
56974.174& 12.6368& 0.101455 \\
56975.174& 12.5512& 0.160767 \\
56976.184& 12.8272& 0.137709 \\
56978.177& 12.4271& 0.117361 \\
56979.175& 12.4385& 0.110956 \\
56980.172& 12.5612& 0.125104 \\
56981.169& 12.4201& 0.128426 \\
56982.166& 12.5571& 0.112808 \\
56983.163& 12.5241& 0.127495 \\
56984.160& 12.4284& 0.0935889\\
56999.121& 13.0574& 0.232812 \\
57013.083& 13.0503& 0.157684 \\
56898.394& 16.4276& 0.155995 \\
56899.395& 16.5029& 0.245028 \\
56900.422& 16.3084& 0.292189 \\
56903.380& 15.8189& 0.162293 \\
56904.354& 16.0975& 0.481578 \\
56906.363& 15.5928& 0.227986 \\
56907.379& 15.7337& 0.342789 \\
56908.366& 15.6151& 0.180709 \\
56909.360& 15.1909& 0.30776  \\
56910.369& 15.0948& 0.136874 \\
56911.348& 14.8637& 0.152151 \\
56959.379& 13.4723& 0.244766 \\
56961.380& 13.5448& 0.18189  \\
56964.338& 13.268 &0.095813  \\
56965.338& 12.9671& 0.103183 \\
56967.330& 13.1403& 0.127211 \\
56969.326& 12.7035& 0.139011 \\
56991.264& 12.8366& 0.0982572\\
57008.240& 13.4449& 0.0784246\\
57010.241& 13.4736& 0.106579 \\
57011.237& 13.5716& 0.142467 \\
57032.176& 13.3936& 0.154944 \\
57035.177& 13.5697& 0.15817  \\
57038.172& 14.2091& 0.192206 \\
57047.108& 14.6907& 0.0955832\\
57050.073& 14.9109& 0.339064 \\
57056.117& 14.5971& 0.267836 \\
57057.070& 14.5593& 0.145233 \\
57058.070& 14.4118& 0.0502472\\
57059.069& 14.4694& 0.130669 \\
57060.069& 14.5013& 0.132637 \\
57065.069& 14.4286& 0.145172 \\
57066.066& 14.3239& 0.101617 \\
57068.065& 14.4446& 0.0949995\\
57069.064& 14.5274& 0.121463 \\
57070.064& 14.5247& 0.0796847\\
57071.063& 14.3549& 0.0858158\\
57074.037& 14.4823& 0.16363  \\
57077.034& 14.0762& 0.127564 \\
57086.003& 14.2022& 0.133571 \\
57091.003& 14.2806& 0.105551 \\
57093.997& 14.616 &0.131274  \\
\end{supertabular}             
      
\end{appendix}                 
                               
\end{document}